\pdfoutput=1
\documentclass[amsmath,amssymb,aps,twocolumn,prb,floatfix,showpacs,10pt]{revtex4-1}
\usepackage{graphicx}
\usepackage{bm}
\usepackage[usenames,dvipsnames]{color}

\begin{document}
\title{Direct molecular dynamics simulation of electrocaloric effect in BaTiO$_3$}
\author{Takeshi Nishimatsu$^{1}$}
\author{Jordan Barr$^{2}$}
\author{Scott P. Beckman$^{2}$}
\affiliation{$^{1}$Institute for Materials Research (IMR), Tohoku University, Sendai 980-8577, Japan\\
  $^{2}$Department of Materials Science and Engineering, Iowa State University, Ames, IA 50011}

\begin{abstract}
The electrocaloric effect (ECE) in BaTiO$_3$ is simulated using two different first-principles based effective Hamiltonian molecular dynamics methods.
The calculations are performed for a wide range of temperatures
(30--900~K) and external electric fields (0--500~kV/cm).
As expected, a large adiabatic temperature change, $\Delta T$, at the
Curie temperature, $T_{\rm C}$, is observed.
It is found that for single crystals of pure BaTiO$_3$,
the temperature range where a large $\Delta T$ is observed is narrow
for small external electric fields ($<50$~kV/cm).
Large fields ($>100$~kV/cm) may be required to broaden the effective
temperature range.
The effect of crystal anisotropy on the ECE $\Delta T$ is also investigated.
It is found that applying an external electric field along the $[001]$ direction
has a larger ECE than those along the $[110]$ and $[111]$ directions.
\end{abstract}



\maketitle

\section{Introduction}
The electrocaloric effect (ECE) is an adiabatic change in the temperature, $\Delta T$,
of a material upon applying an external electric field.
In particular, if an electric field is applied to a ferroelectric material
at just above its phase transition temperature, $T_{\rm C}$, and the field is then removed,
a large reduction in temperature is expected.
It is widely believed that this effect is applicable
to solid-state refrigeration technologies.

In addition, recent developments in the techniques of vapor deposition
enable the production of
defect-free single-crystal ferroelectric thin films.
Such high-quality films allow for the application of large external electric fields,
which cannot be applied to bulk polycrystalline specimens.
Consequently, this advance in processing allows for investigation of the ECE in these ferroelectric thin films.
There has been much interest in
this subject and experimental studies have shown the possibility of
creating materials with a relatively large electrocaloric
response\cite{mischenko2006,mischenko:APL:242912,bai:APL96:192902}.

There have been several computational simulations of the ECE published in the literature.
Ponomareva and Lisenkov have investigated the ECE of Ba$_{0.5}$Sr$_{0.5}$TiO$_3$
using Monte Carlo methods\cite{PhysRevLett.108.167604:Ponomareva:ECE}.
Rose and Cohen\cite{PhysRevLett.109.187604:Cohen:ECE} have used molecular
dynamics (MD) simulations and core-shell interatomic potentials to model the ECE in bulk
LiNbO$_3$. Also using this form of atomic potential, Chen and
Fang\cite{Chen:Fang:PhysicaB:v415:y2013:p14} have simulated the ECE in
BaTiO$_3$ nanoparticles.
We have also used the so-called \textit{indirect} MD method, discussed below,
to calculate the ECE of bulk BaTiO$_3$ using a first-principles
based effective Hamiltonian\cite{Scott:ECE}.
All of these simulations find a large ECE is observed just above $T_{\rm C}$
due to the large change in entropy when transforming from the paraelectric to ferroelectric phase.

Following this introduction we present the methods used
to calculate the ECE and in particular we introduce two
MD methods that can be used to study the electro-thermal coupling.
The results are presented in Sec.~\ref{sec:results} and in
Sec.~\ref{sec:summary}, the paper is summarized and conclusions are given.
The methods and results presented here are a
full detailed review that extends and explains the preliminary results
given in the proceedings
Ref.~\onlinecite{Jordan:Takeshi:Scott:2013:MRS}.

\section{Methods of calculation and formalism}
\label{sec:Formalism}
\subsection{Effective Hamiltonian}
\label{subsec:effective:Hamiltonian}

\begin{figure}
  \centering
  \includegraphics[width=80mm]{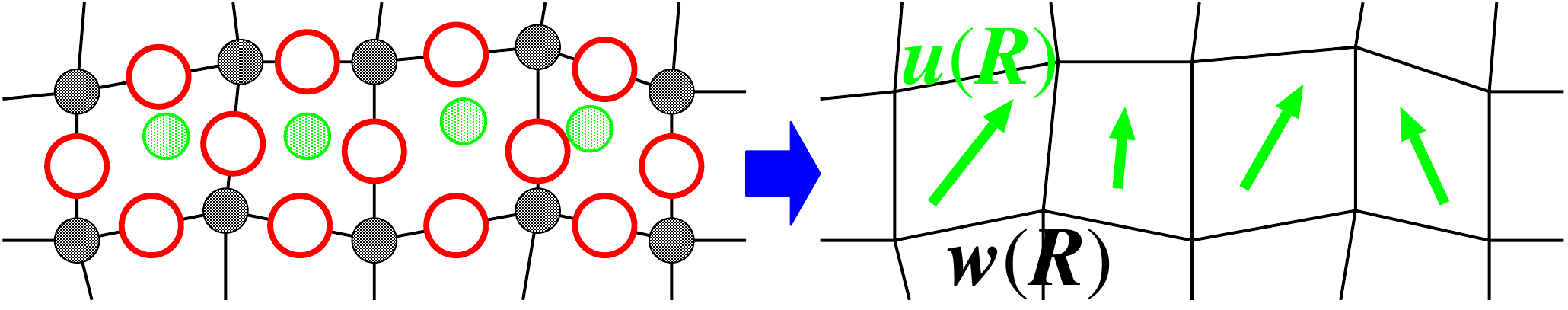}
  \caption{(Color online) Schematic illustration of coarse graining utilized in the
    effective Hamiltonian (\ref{eq:Effective:Hamiltonian}).
    The number of degrees of freedom per unit cell
    is reduced from 15 (5 atoms $\times$ 3 $xyz$-directions)
    to 6 (two 3-dimensional vectors).}
  \label{fig:CoarseGraining}
\end{figure}

The effective Hamiltonian, constructed from first-principles calculations, and
used in the present MD simulations is essentially the same as that in
Ref.~\onlinecite{Waghmare:R:1997PRB,Nishimatsu:feram:PRB2008}
\begin{multline}
  \label{eq:Effective:Hamiltonian}
  H^{\rm eff}
  = \frac{M^*_{\rm dipole}}{2} \sum_{\bm{R},\alpha}\dot{u}_\alpha^2(\bm{R})
  + \frac{M^*_{\rm acoustic}}{2}\sum_{\bm{R},\alpha}\dot{w}_\alpha^2(\bm{R})\\
  + V^{\rm self}(\{\bm{u}\})+V^{\rm dpl}(\{\bm{u}\})+V^{\rm short}(\{\bm{u}\})\\
  + V^{\rm elas,\,homo}(\eta_1,\dots,\eta_6)+V^{\rm elas,\,inho}(\{\bm{w}\})\\
  + V^{\rm coup,\,homo}(\{\bm{u}\}, \eta_1,\dots,\eta_6)+V^{\rm coup,\,inho}(\{\bm{u}\}, \{\bm{w}\})\\
  -Z^*\sum_{\bm{R}}\bm{\mathcal{E}}\!\cdot\!\bm{u}(\bm{R})~.
\end{multline}
The true atomic structure has properties determined by the
complex chemical bonding between the atoms,
but in the model system the complexity is
reduced; the collective atomic motion is
coarse-grained by local soft mode vectors, $\bm{u}(\bm{R})$, and
local acoustic displacement vectors, $\bm{w}(\bm{R})$,
of each unit cell located at $\bm{R}$ in a simulation supercell as depicted in Fig.~\ref{fig:CoarseGraining}.
Each term in the Hamiltonian bears a physical significance:
$\frac{M^*_{\rm dipole}}{2} \sum_{\bm{R},\alpha}\dot{u}_\alpha^2(\bm{R})$
is the kinetic energy of local soft modes with effective mass $M^*_{\rm dipole}$,
$\frac{M^*_{\rm acoustic}}{2}\sum_{\bm{R},\alpha}\dot{w}_\alpha^2(\bm{R})$
is the kinetic energy of acoustic displacements with effective mass $M^*_{\rm acoustic}$,
$V^{\rm self}(\{\bm{u}\})$ is the local mode self energy,
$V^{\rm dpl}(\{\bm{u}\})$ is the long-ranged dipole-dipole interaction,
$V^{\rm short}(\{\bm{u}\})$ is the short-ranged interaction between local soft modes,
$V^{\rm elas,\,homo}(\eta_1,\dots,\eta_6)$
is the elastic energy from homogeneous strains $\eta_1,\dots,\eta_6$ (Voigt notation; $\eta_1=e_{11}$, $\eta_4=e_{23}$),
$V^{\rm elas,\,inho}(\{\bm{w}\})$ is the elastic energy from inhomogeneous strains,
$V^{\rm coup,\,homo}(\{\bm{u}\}, \eta_1,\dots,\eta_6)$
is the coupling between the local soft modes and the homogeneous strain,
$V^{\rm coup,\,inho}(\{\bm{u}\}, \{\bm{w}\})$
is the coupling between the soft modes and the inhomogeneous strains, and
$-Z^*\sum_{\bm{R}}\bm{\mathcal{E}}\!\cdot\!\bm{u}(\bm{R})$
is the interaction between dipoles $Z^*\bm{u}(\bm{R})$ and external electric field $\mathcal{E}$,
here $Z^*$ is the effective charge of the soft mode per unit cell.
The coarse graining and calculation of $V^{\rm dpl}(\{\bm{u}\})$ using fast-Fourier transform (FFT)
enabled us to perform fast MD simulations in a large supercell,
and was previously applied to bulk relaxor ferroelectrics~\cite{Waghmare:C:B:2003,Burton:C:W:PRB:72:p064113:2005}.
A more detailed explanation of the effective Hamiltonian
can be found in
Refs.~\onlinecite{King-Smith:V:1994},
\onlinecite{Zhong:V:R:PRB:v52:p6301:1995},
\onlinecite{Nishimatsu:feram:PRB2008}, and
\onlinecite{Nishimatsu.PhysRevB.82.134106}.
The set of parameters for the effective Hamiltonian for BaTiO$_3$ are
listed in Ref.~\onlinecite{Nishimatsu.PhysRevB.82.134106}.
A temperature dependent effective negative pressure given by $p=-0.005T$~GPa is
applied to simulate the thermal expansion~\cite{Nishimatsu.PhysRevB.82.134106}.

\subsection{MD methods and conditions}
\label{subsec:MD:conditions}

MD simulations for BaTiO$_3$
using the effective Hamiltonian in Eq.~(\ref{eq:Effective:Hamiltonian})
are performed with the \texttt{feram} software.
\texttt{feram} is distributed freely under the GNU General Public License (GPL) and can be found at (\url{http://loto.sourceforge.net/feram/}). Details of the code can be found in Ref.~\onlinecite{Nishimatsu:feram:PRB2008}. For the calculations presented here version \texttt{feram-0.21.02} was used.
Examples of the
input files are packaged within the source code under
the \texttt{feram-X.YY.ZZ/src/26example-BaTiO3-acoustic-MD/} directory.

As described in Ref.~\onlinecite{Scott:ECE},
for the so-called \textit{indirect} approach,
the pyroelectric coefficient is predicted as a function of
applied electric field and temperature.
Then, using the thermodynamic relationship
\begin{equation}
  \label{eq:DeltaT}
  \Delta T = T\int^{\mathcal{E}_2}_{\mathcal{E}_1}\frac{1}{C(\mathcal{E},T)}
  \left.\left(\frac{\partial P}{\partial T}\right)\right|_T d{\mathcal{E}}\ ,
\end{equation}
the ECE $\Delta T$ is related to the temperature and external electric field.
Here, $P$ is polarization, and the external electric field, $\mathcal{E}$, is switched
from $\mathcal{E}_1$ to $\mathcal{E}_2$.
The specific heat capacity, $C$, is approximated as a constant and
its value is taken from experimental observation~\cite{Yi2004135}.

However, it is well known that $C$ is dependent on the
external electric field and temperature,
especially near the phase transition temperature $T_{\rm C}$.
Moreover, under a weak external electric field, it is
numerically difficult to determine ${\partial P}/{\partial T}$ at $T_{\rm C}$.
Therefore, a better approach involves the
so-called \textit{direct} method to estimate the ECE $\Delta T$.
The procedure involves two steps:
first constant-temperature MD is performed for a fixed external electric field, $\mathcal{E}$,
in the canonical ensemble using the velocity-scaling thermostat.
This allows the system to equilibriate.
Next, the external electric field is switched off and the system is simultaneously changed to a
constant-energy MD in the microcanonical ensemble that is allowed to evolve using the
leapfrog method. The final state at the end of the constant-temperature MD is used as the
initial state of the constant energy MD.
A time step of $\Delta t = 2$~fs is used in both ensembles.

There are two ways to determine the acoustic displacements, $\{\bm{w}(\bm{R})\}$, at each
time step of the MD simulation.
One allows for the natural time evolution of the effective
Hamiltonian Eq.~(\ref{eq:Effective:Hamiltonian}) with MD.
As discussed in Sec.~\ref{subsec:effective:Hamiltonian},
the number of degrees of freedom is reduced from
15 (5 atoms $\times$ 3 $xyz$-directions) to 6
(two 3-dimensional vectors) and consequently the
specific heat capacity implicitly becomes $6/15=2/5$ of the real system.
Another approach is to optimize $\{\bm{w}(\bm{R})\}$ such that
$V^{\rm elas,\,inho}(\{\bm{w}\})+V^{\rm coup,\,inho}(\{\bm{u}\}, \{\bm{w}\})$
is minimized at each time step according to $\{\bm{u}(\bm{R})\}$.
In this case, $\{\bm{w}(\bm{R})\}$ is fully dependent on $\{\bm{u}(\bm{R})\}$,
and the degrees of freedom is further reduced from 6 to 3, meaning that the
specific heat capacity of the calculated system is $1/5$ of the real system.
In both of these \textit{direct} MD methods, $\{\bm{u}(\bm{R})\}$ evolves normally
according to the effective Hamiltonian of Eq.~(\ref{eq:Effective:Hamiltonian}).

\begin{figure}
  \centering
  \includegraphics[width=80mm]{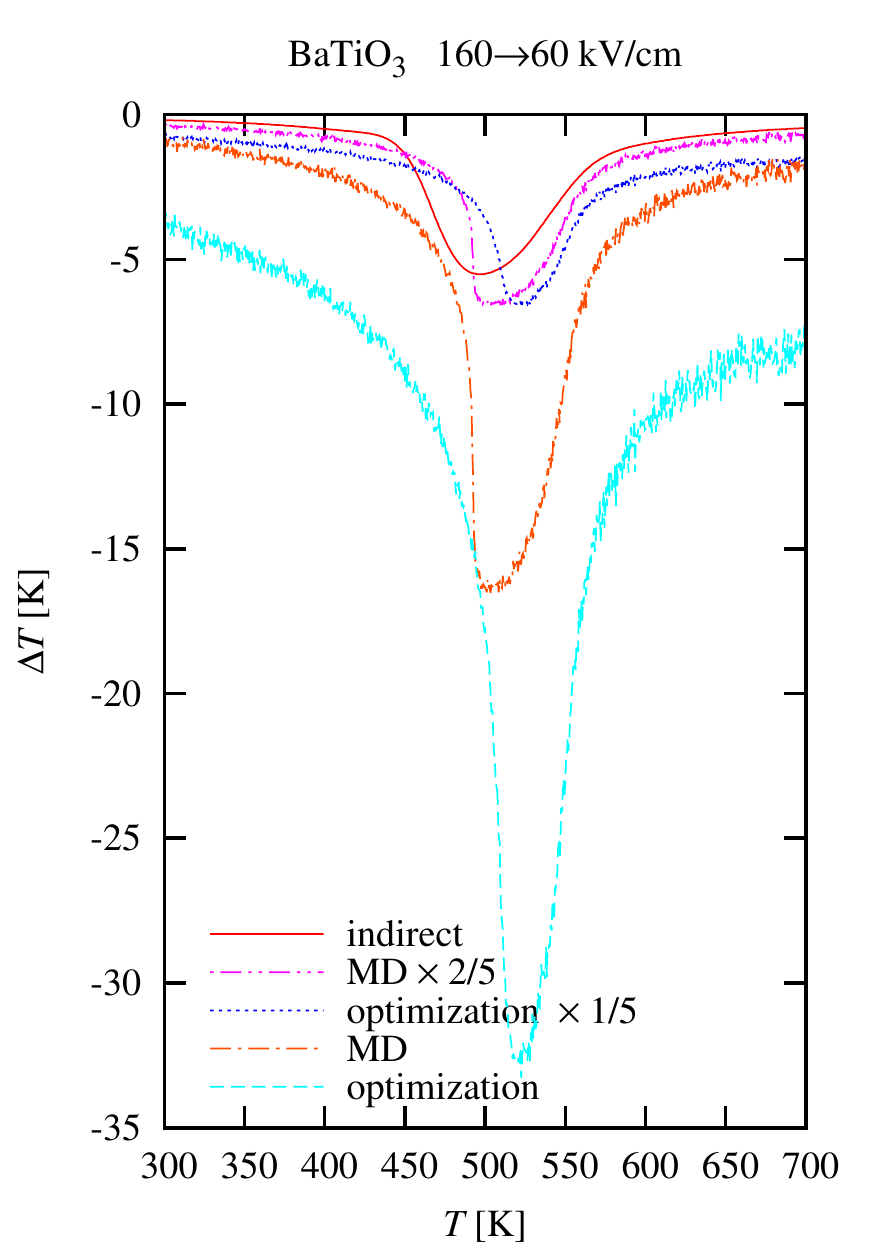}
  \caption{(Color online) A comparison of the three MD methods to simulate the ECE.
    The temperature dependence of the ECE $\Delta T$ of BaTiO$_3$ is plotted versus
    ambient temperature for switching the applied external electric field from
    160 to 60~kV/cm in the $[001]$ direction.
    The solid (red) line is the data from the
                     \textit{indirect} method described in Ref.~\onlinecite{Scott:ECE}.
    The dashed (cyan) line is the raw data from the
                     \textit{direct} method with optimization of the acoustic displacements $\bm{w}(\bm{k})$ and
                     the dotted (blue) line is the data scaled by $1/5$.
    The chain (orange) line is raw data of the \textit{direct} method
                     in which acoustic displacements are treated by MD and
                     the double-dotted-chain (magenta) line is the data scaled by $2/5$.}
  \label{fig:compare2}
\end{figure}

The ECE response is calculated using the \textit{indirect} method, \textit{direct MD} method, and
\textit{direct optimized} method and are compared in Fig.~\ref{fig:compare2}.
The values of $\Delta T$, corrected for the underestimated heat capacity in
the \textit{direct} methods, are also given in the figure.
The differing results for these three methods come primarily from two effects.
One is that in the \textit{direct} methods, the effective negative pressure, $p=-0.005T$~GPa,
is kept constant at the starting temperature,
even though the constant energy evolution of the system causes a reduction in temperature.
Another reason for the difference between the results is
that the temperature and external electric field dependence of the specific heat capacity is
automatically included in the \textit{direct} methods whereas it must be approximated from
experiment in the \textit{indirect} methods.

The \textit{direct MD} approach gives half of the raw temperature
change $\Delta T$ during the constant-energy MD simulation as
compared to the \textit{direct optimized} approach due to the
difference in the degrees of freedom. However, error is introduced because
the $T$-\textit{dependent} effective negative pressure is held
constant during the constant energy temperature change.
Therefore, in principle, the \textit{direct MD} approach may give more accurate results
than the \textit{direct optimized} approach.
In the study presented here, however, the \textit{direct optimized} method is used
because it is shown in Fig.~\ref{fig:compare2} to produce nearly equivalent results
and is almost twice as efficient computationally.

\begin{figure}
  \centering
  \includegraphics[width=72mm]{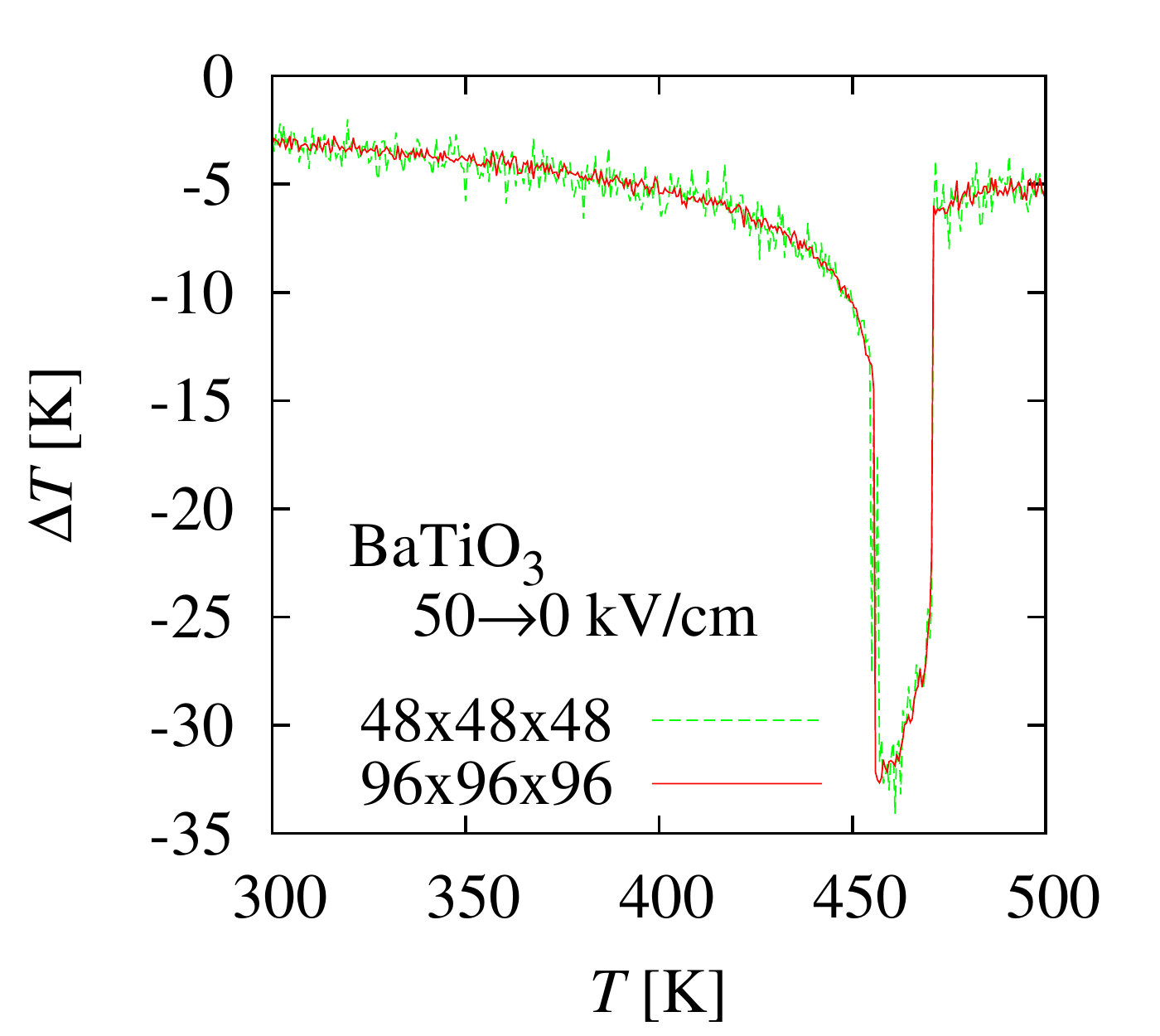}
  \caption{(Color online) System size dependence of MD simulations of ECE.
    Dashed (green) line is calculated with a $L_x\times L_y\times L_z = 48\times 48\times 48$ super cell
    and has larger fluctuations of $\Delta T$ than
    the solid (red) line of super cell size $96\times 96\times 96$.
    In both simulations, the external electric field is altered from 50 to 0~kV/cm.}
  \label{fig:c48vs96}
\end{figure}

In Fig.~\ref{fig:c48vs96}, the $\Delta T$ is compared for two supercells of sizes
$L_x\times L_y\times L_z = 48 \times 48 \times 48$ and
$96 \times 96 \times 96$ unit cells.
The greater fluctuations in the $\Delta T$ in the $48 \times 48 \times 48$ system is likely due to the
system-wide thermal fluctuations that evolve in the constant-energy MD.
Therefore, a system size of $96 \times 96 \times 96$ is employed for the simulations presented here.
To accurately capture the temperature dependent phase transformations and account for the
thermal fluctuations discussed above,
a temperature step size of 0.5~K is used
and numerous simulations are performed, e.g. 1,400 sets of
MD simulations are performed for a sweep from 300.0 to 999.5~K.

\section{Results and Discussion}
\label{sec:results}

\begin{figure}
  \centering
  \includegraphics[width=82mm]{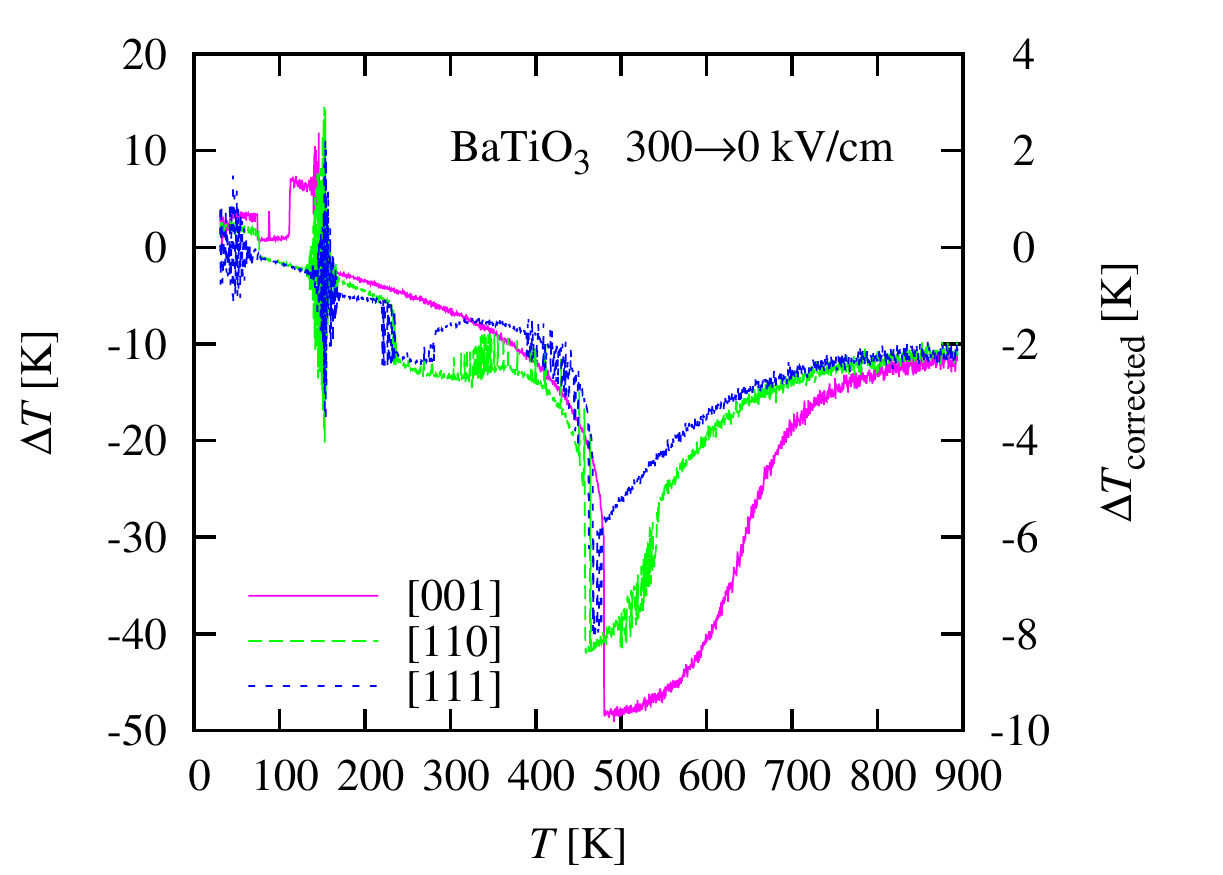}
  \caption{(Color online) The effect of crystal anisotropy on the ECE $\Delta T$.
    The strength of the initial applied field is
    $|\mathcal{E}|=300$~kV/cm and the final field is $0$~kV/cm for all datasets.
    The direction of the applied fields differ:
    the solid  (magenta) line is for $[001]$,
    the dashed (green)   line is for $[110]$, and
    the dotted (blue)    line is for $[111]$.
    The left ordinate is the
    raw data $\Delta T$ for the direct method and the right ordinate is the
    corrected data $\Delta T_{\rm corrected}=\frac{1}{5}\Delta T$ that
    accounts for the underestimated specific heat capacity.
    }
  \label{fig:anisotropy}
\end{figure}

Using this method the impact of crystallographic anisotropy on the ECE is
investigated by examining $\Delta T$ as a function of temperature for
electric fields applied along the  $[001]$, $[110]$, and $[111]$ directions.
As seen in Fig.~\ref{fig:anisotropy} there exist singularities at the temperatures
associated with the
tetragonal-to-orthorhombic and
orthorhombic-to-rhombohedral transformations; however,
$|\Delta T|$ is maximum at the cubic-to-tetragonal transition temperature $T_{\rm C}$
for all applied fields.
It is clearly seen that when the field is applied in the $[001]$ direction $\Delta T$ is the
largest and the available temperature range for the large ECE is the greatest.
This is because of the strong coupling between the external electric
field and the internal dipole moment. The first ferroelectric phase at
$T<T_{\rm C}$ is tetragonal and therefore the applied electric field
acts to broaden the transformation, effectively increasing the transformation
temperature.
This is of practical importance for solid state cooling technologies because the engineering
devices must be tuned to operate within a specific range of temperatures.
For these applications ECE research should be focused on the cases where the applied
electric field aligns with the intrinsic polarization at the Curie temperature, which in the
case of BaTiO$_3$ is the $[001]$ direction.

\begin{figure*}
  \centering
  \includegraphics[height=152mm,angle=-90]{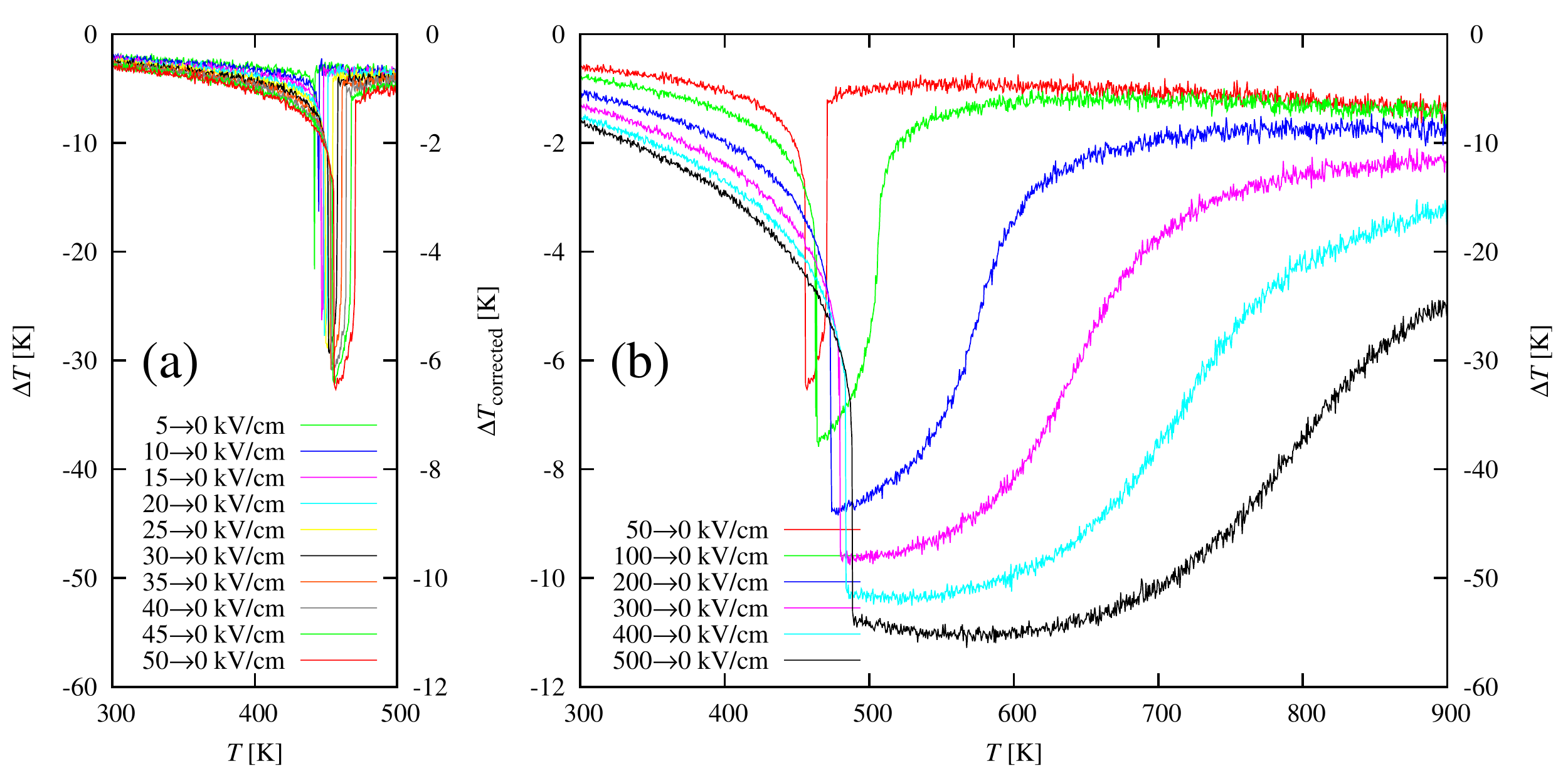}
  \caption{(Color online) The temperature dependence of $\Delta T$ for
  various initial external electric fields along the $[001]$ direction. In frame
    (a) the external electric field switches from $\mathcal{E}_z=$~5--50 to 0~kV/cm and in
    frame (b) it switches from $\mathcal{E}_z=$~50--500 to 0~kV/cm.
    There is an ordinate for both the
    raw data $\Delta T$ and for the
    corrected data $\Delta T_{\rm corrected}=\frac{1}{5}\Delta T$ that
    accounts for the underestimated specific heat capacity.}
  \label{fig:clow}
\end{figure*}

In Fig.~\ref{fig:clow}, the temperature dependence for the ECE $\Delta T$ of BaTiO$_3$,
under various initial external electric fields is compared.
It can be seen in Fig.~\ref{fig:clow}(a),
that even with a small initial external electric field ($<50$~kV/cm),
BaTiO$_3$ gives a large $\Delta T$,
but the temperature range where this large $\Delta T$ can be obtained is narrow.
This result is consistent with previous theoretical studies of
other ferroelectric materials\cite{PhysRevLett.108.167604:Ponomareva:ECE,PhysRevLett.109.187604:Cohen:ECE,Chen:Fang:PhysicaB:v415:y2013:p14}
in which $|\Delta T|$ has a peak at $T_{\rm C}$.
By increasing the applied fields ($>100$~kV/cm) the range of applicable temperatures
broadens as shown in Fig.~\ref{fig:clow}(b).
This is again due to the strong coupling between the tetragonal
polarization and the external electric field, which both broadens
the thermal range of the transformation and increases the
pyroelectric response.
For engineering applications it may be necessary to apply as large of field as possible
to allow for an operating temperature range that is useful.

\section{Summary and Conclusions}
\label{sec:summary}

In this work we have described two MD \textit{direct} methods to evaluate the ECE
and they are compared to the previous \textit{indirect} approach.
The reduction in the degrees of freedom associated with this effective Hamiltonian
method are explained and its impact on the specific heat capacity, as applied to the
\textit{direct MD} and \textit{direct optimized} methods, are observed.
The three methods demonstrated here are found to produce roughly equivalent
results; however, based on the computational errors discussed above it is
concluded that the \textit{direct MD} method is in principle more accurate although it is
slower than the \textit{direct optimized} method used in these calculations.

The effect of crystal anisotropy is found to be great.
For BaTiO$_3$, applying a field in the $[001]$ direction results in the greatest electrocaloric response.
This is good news for ferroelectric thin-films grown by vapor deposition methods
that normally have the $[001]$ polarization direction perpendicular to the film.
It is also found that increasing the electric field that is switched broadens the range of
applicable temperatures where a large ECE can be achieved.

These observations are physically intuitive.
The applied electric field acts to elevate the
transformation temperature so that for $T>T_{\rm C}$ the crystal remains in the
ordered ferroelectric phase.
When the field is removed the crystal transforms to the disordered,
paraelectric state and the change in the entropy results in an adiabatic change
in temperature.
The magnitude of the temperature change is directly proportional to
the magnitude of the change in electric field.
Although applying an electric field in arbitrary directions does act to stabilize the
ordered phase above $T_{\rm C}$,
the largest $\Delta T$ and widest range of applicable temperatures is achieved
when the external field is applied parallel to the polarization direction of the first
ferroelectric phase just below the Curie temperature.
In the case of BaTiO$_3$ this is the $[001]$ direction, but for other
ferroelectric crystals the direction of applied field should be taken from
knowledge of their phase diagrams.
Because the electric field acts to raise the transformation temperature, having a larger
switching field naturally broadens the range of applicable temperatures.


\section*{Acknowledgments}
Computational resources
were provided by the Center for Computational Materials Science,
Institute for Materials Research (CCMS-IMR), Tohoku University.
We thank the staff at CCMS-IMR for their constant effort.
This work was supported in part by JSPS KAKENHI Grant Number 25400314.
The work of S.P.B. and J.A.B. was supported by the U.S.
National Science Foundation (NSF) through grant DMR-1105641.

\bibliography{biblio/ferroelectrics}

\begin{thebibliography}{16}%
\makeatletter
\providecommand \@ifxundefined [1]{%
 \@ifx{#1\undefined}
}%
\providecommand \@ifnum [1]{%
 \ifnum #1\expandafter \@firstoftwo
 \else \expandafter \@secondoftwo
 \fi
}%
\providecommand \@ifx [1]{%
 \ifx #1\expandafter \@firstoftwo
 \else \expandafter \@secondoftwo
 \fi
}%
\providecommand \natexlab [1]{#1}%
\providecommand \enquote  [1]{``#1''}%
\providecommand \bibnamefont  [1]{#1}%
\providecommand \bibfnamefont [1]{#1}%
\providecommand \citenamefont [1]{#1}%
\providecommand \href@noop [0]{\@secondoftwo}%
\providecommand \href [0]{\begingroup \@sanitize@url \@href}%
\providecommand \@href[1]{\@@startlink{#1}\@@href}%
\providecommand \@@href[1]{\endgroup#1\@@endlink}%
\providecommand \@sanitize@url [0]{\catcode `\\12\catcode `\$12\catcode
  `\&12\catcode `\#12\catcode `\^12\catcode `\_12\catcode `\%12\relax}%
\providecommand \@@startlink[1]{}%
\providecommand \@@endlink[0]{}%
\providecommand \url  [0]{\begingroup\@sanitize@url \@url }%
\providecommand \@url [1]{\endgroup\@href {#1}{\urlprefix }}%
\providecommand \urlprefix  [0]{URL }%
\providecommand \Eprint [0]{\href }%
\providecommand \doibase [0]{http://dx.doi.org/}%
\providecommand \selectlanguage [0]{\@gobble}%
\providecommand \bibinfo  [0]{\@secondoftwo}%
\providecommand \bibfield  [0]{\@secondoftwo}%
\providecommand \translation [1]{[#1]}%
\providecommand \BibitemOpen [0]{}%
\providecommand \bibitemStop [0]{}%
\providecommand \bibitemNoStop [0]{.\EOS\space}%
\providecommand \EOS [0]{\spacefactor3000\relax}%
\providecommand \BibitemShut  [1]{\csname bibitem#1\endcsname}%
\let\auto@bib@innerbib\@empty
\bibitem [{\citenamefont {Mischenko}\ \emph
  {et~al.}(2006{\natexlab{a}})\citenamefont {Mischenko}, \citenamefont {Zhang},
  \citenamefont {Scott}, \citenamefont {Whatmore},\ and\ \citenamefont
  {Mathur}}]{mischenko2006}%
  \BibitemOpen
  \bibfield  {author} {\bibinfo {author} {\bibfnamefont {A.}~\bibnamefont
  {Mischenko}}, \bibinfo {author} {\bibfnamefont {Q.}~\bibnamefont {Zhang}},
  \bibinfo {author} {\bibfnamefont {J.}~\bibnamefont {Scott}}, \bibinfo
  {author} {\bibfnamefont {R.}~\bibnamefont {Whatmore}}, \ and\ \bibinfo
  {author} {\bibfnamefont {N.}~\bibnamefont {Mathur}},\ }\href@noop {}
  {\bibfield  {journal} {\bibinfo  {journal} {Science}\ }\textbf {\bibinfo
  {volume} {311}},\ \bibinfo {pages} {1270} (\bibinfo {year}
  {2006}{\natexlab{a}})}\BibitemShut {NoStop}%
\bibitem [{\citenamefont {Mischenko}\ \emph
  {et~al.}(2006{\natexlab{b}})\citenamefont {Mischenko}, \citenamefont {Zhang},
  \citenamefont {Whatmore}, \citenamefont {Scott},\ and\ \citenamefont
  {Mathur}}]{mischenko:APL:242912}%
  \BibitemOpen
  \bibfield  {author} {\bibinfo {author} {\bibfnamefont {A.~S.}\ \bibnamefont
  {Mischenko}}, \bibinfo {author} {\bibfnamefont {Q.}~\bibnamefont {Zhang}},
  \bibinfo {author} {\bibfnamefont {R.~W.}\ \bibnamefont {Whatmore}}, \bibinfo
  {author} {\bibfnamefont {J.~F.}\ \bibnamefont {Scott}}, \ and\ \bibinfo
  {author} {\bibfnamefont {N.~D.}\ \bibnamefont {Mathur}},\ }\href {\doibase
  10.1063/1.2405889} {\bibfield  {journal} {\bibinfo  {journal} {Applied
  Physics Letters}\ }\textbf {\bibinfo {volume} {89}},\ \bibinfo {eid} {242912}
  (\bibinfo {year} {2006}{\natexlab{b}})}\BibitemShut {NoStop}%
\bibitem [{\citenamefont {Bai}\ \emph {et~al.}(2010)\citenamefont {Bai},
  \citenamefont {Zheng},\ and\ \citenamefont {Shi}}]{bai:APL96:192902}%
  \BibitemOpen
  \bibfield  {author} {\bibinfo {author} {\bibfnamefont {Y.}~\bibnamefont
  {Bai}}, \bibinfo {author} {\bibfnamefont {G.}~\bibnamefont {Zheng}}, \ and\
  \bibinfo {author} {\bibfnamefont {S.}~\bibnamefont {Shi}},\ }\href {\doibase
  10.1063/1.3430045} {\bibfield  {journal} {\bibinfo  {journal} {Applied
  Physics Letters}\ }\textbf {\bibinfo {volume} {96}},\ \bibinfo {eid} {192902}
  (\bibinfo {year} {2010})}\BibitemShut {NoStop}%
\bibitem [{\citenamefont {Ponomareva}\ and\ \citenamefont
  {Lisenkov}(2012)}]{PhysRevLett.108.167604:Ponomareva:ECE}%
  \BibitemOpen
  \bibfield  {author} {\bibinfo {author} {\bibfnamefont {I.}~\bibnamefont
  {Ponomareva}}\ and\ \bibinfo {author} {\bibfnamefont {S.}~\bibnamefont
  {Lisenkov}},\ }\href {\doibase 10.1103/PhysRevLett.108.167604} {\bibfield
  {journal} {\bibinfo  {journal} {Phys. Rev. Lett.}\ }\textbf {\bibinfo
  {volume} {108}},\ \bibinfo {pages} {167604} (\bibinfo {year}
  {2012})}\BibitemShut {NoStop}%
\bibitem [{\citenamefont {Rose}\ and\ \citenamefont
  {Cohen}(2012)}]{PhysRevLett.109.187604:Cohen:ECE}%
  \BibitemOpen
  \bibfield  {author} {\bibinfo {author} {\bibfnamefont {M.~C.}\ \bibnamefont
  {Rose}}\ and\ \bibinfo {author} {\bibfnamefont {R.~E.}\ \bibnamefont
  {Cohen}},\ }\href {\doibase 10.1103/PhysRevLett.109.187604} {\bibfield
  {journal} {\bibinfo  {journal} {Phys. Rev. Lett.}\ }\textbf {\bibinfo
  {volume} {109}},\ \bibinfo {pages} {187604} (\bibinfo {year}
  {2012})}\BibitemShut {NoStop}%
\bibitem [{\citenamefont {Chen}\ and\ \citenamefont
  {Fang}(2013)}]{Chen:Fang:PhysicaB:v415:y2013:p14}%
  \BibitemOpen
  \bibfield  {author} {\bibinfo {author} {\bibfnamefont {X.}~\bibnamefont
  {Chen}}\ and\ \bibinfo {author} {\bibfnamefont {C.}~\bibnamefont {Fang}},\
  }\href {\doibase http://dx.doi.org/10.1016/j.physb.2013.01.033} {\bibfield
  {journal} {\bibinfo  {journal} {Physica B: Condensed Matter}\ }\textbf
  {\bibinfo {volume} {415}},\ \bibinfo {pages} {14 } (\bibinfo {year}
  {2013})}\BibitemShut {NoStop}%
\bibitem [{\citenamefont {Beckman}\ \emph {et~al.}(2012)\citenamefont
  {Beckman}, \citenamefont {Wan}, \citenamefont {Barr},\ and\ \citenamefont
  {Nishimatsu}}]{Scott:ECE}%
  \BibitemOpen
  \bibfield  {author} {\bibinfo {author} {\bibfnamefont {S.~P.}\ \bibnamefont
  {Beckman}}, \bibinfo {author} {\bibfnamefont {L.~F.}\ \bibnamefont {Wan}},
  \bibinfo {author} {\bibfnamefont {J.~A.}\ \bibnamefont {Barr}}, \ and\
  \bibinfo {author} {\bibfnamefont {T.}~\bibnamefont {Nishimatsu}},\
  }\href@noop {} {\bibfield  {journal} {\bibinfo  {journal} {Materials
  Letters}\ }\textbf {\bibinfo {volume} {89}},\ \bibinfo {pages} {254}
  (\bibinfo {year} {2012})}\BibitemShut {NoStop}%
\bibitem [{\citenamefont {Barr}\ \emph {et~al.}(2013)\citenamefont {Barr},
  \citenamefont {Beckman},\ and\ \citenamefont
  {Nishimatsu}}]{Jordan:Takeshi:Scott:2013:MRS}%
  \BibitemOpen
  \bibfield  {author} {\bibinfo {author} {\bibfnamefont {J.~A.}\ \bibnamefont
  {Barr}}, \bibinfo {author} {\bibfnamefont {S.~P.}\ \bibnamefont {Beckman}}, \
  and\ \bibinfo {author} {\bibfnamefont {T.}~\bibnamefont {Nishimatsu}},\ }in\
  \href@noop {} {\emph {\bibinfo {booktitle} {Nanoscale Thermoelectric
  Materials, Thermal and Electrical Transport, and Applications to Solid-State
  Cooling and Power Generation}}},\ \bibinfo {series and number} {MRS Spring
  Meeting Proceedings},\ \bibinfo {editor} {edited by\ \bibinfo {editor}
  {\bibfnamefont {S.}~\bibnamefont {Beckman}}, \bibinfo {editor} {\bibfnamefont
  {H.}~\bibnamefont {B\"ottner}}, \bibinfo {editor} {\bibfnamefont
  {Y.}~\bibnamefont {Chalopin}}, \bibinfo {editor} {\bibfnamefont
  {C.}~\bibnamefont {Dames}}, \bibinfo {editor} {\bibfnamefont {P.~A.}\
  \bibnamefont {Greaney}}, \bibinfo {editor} {\bibfnamefont {P.}~\bibnamefont
  {Hopkins}}, \bibinfo {editor} {\bibfnamefont {B.}~\bibnamefont {Li}},
  \bibinfo {editor} {\bibfnamefont {T.}~\bibnamefont {Mori}}, \bibinfo {editor}
  {\bibfnamefont {T.}~\bibnamefont {Nishimatsu}}, \bibinfo {editor}
  {\bibfnamefont {K.}~\bibnamefont {Pipe}}, \ and\ \bibinfo {editor}
  {\bibfnamefont {R.}~\bibnamefont {Venkatasubramanian}}}\ (\bibinfo {year}
  {2013})\BibitemShut {NoStop}%
\bibitem [{\citenamefont {Waghmare}\ and\ \citenamefont
  {Rabe}(1997)}]{Waghmare:R:1997PRB}%
  \BibitemOpen
  \bibfield  {author} {\bibinfo {author} {\bibfnamefont {U.~V.}\ \bibnamefont
  {Waghmare}}\ and\ \bibinfo {author} {\bibfnamefont {K.~M.}\ \bibnamefont
  {Rabe}},\ }\href@noop {} {\bibfield  {journal} {\bibinfo  {journal} {Phys.
  Rev. B}\ }\textbf {\bibinfo {volume} {55}},\ \bibinfo {pages} {6161}
  (\bibinfo {year} {1997})}\BibitemShut {NoStop}%
\bibitem [{\citenamefont {Nishimatsu}\ \emph {et~al.}(2008)\citenamefont
  {Nishimatsu}, \citenamefont {Waghmare}, \citenamefont {Kawazoe},\ and\
  \citenamefont {Vanderbilt}}]{Nishimatsu:feram:PRB2008}%
  \BibitemOpen
  \bibfield  {author} {\bibinfo {author} {\bibfnamefont {T.}~\bibnamefont
  {Nishimatsu}}, \bibinfo {author} {\bibfnamefont {U.~V.}\ \bibnamefont
  {Waghmare}}, \bibinfo {author} {\bibfnamefont {Y.}~\bibnamefont {Kawazoe}}, \
  and\ \bibinfo {author} {\bibfnamefont {D.}~\bibnamefont {Vanderbilt}},\
  }\href@noop {} {\bibfield  {journal} {\bibinfo  {journal} {Phys. Rev. B}\
  }\textbf {\bibinfo {volume} {78}},\ \bibinfo {pages} {104104} (\bibinfo
  {year} {2008})}\BibitemShut {NoStop}%
\bibitem [{\citenamefont {Waghmare}\ \emph {et~al.}(2003)\citenamefont
  {Waghmare}, \citenamefont {Cockayne},\ and\ \citenamefont
  {Burton}}]{Waghmare:C:B:2003}%
  \BibitemOpen
  \bibfield  {author} {\bibinfo {author} {\bibfnamefont {U.~V.}\ \bibnamefont
  {Waghmare}}, \bibinfo {author} {\bibfnamefont {E.~J.}\ \bibnamefont
  {Cockayne}}, \ and\ \bibinfo {author} {\bibfnamefont {B.~P.}\ \bibnamefont
  {Burton}},\ }\href@noop {} {\bibfield  {journal} {\bibinfo  {journal}
  {Ferroelectrics}\ }\textbf {\bibinfo {volume} {291}},\ \bibinfo {pages} {187}
  (\bibinfo {year} {2003})}\BibitemShut {NoStop}%
\bibitem [{\citenamefont {Burton}\ \emph {et~al.}(2005)\citenamefont {Burton},
  \citenamefont {Cockayne},\ and\ \citenamefont
  {Waghmare}}]{Burton:C:W:PRB:72:p064113:2005}%
  \BibitemOpen
  \bibfield  {author} {\bibinfo {author} {\bibfnamefont {B.~P.}\ \bibnamefont
  {Burton}}, \bibinfo {author} {\bibfnamefont {E.}~\bibnamefont {Cockayne}}, \
  and\ \bibinfo {author} {\bibfnamefont {U.~V.}\ \bibnamefont {Waghmare}},\
  }\href@noop {} {\bibfield  {journal} {\bibinfo  {journal} {Phys. Rev. B}\
  }\textbf {\bibinfo {volume} {72}},\ \bibinfo {pages} {064113} (\bibinfo
  {year} {2005})}\BibitemShut {NoStop}%
\bibitem [{\citenamefont {King-Smith}\ and\ \citenamefont
  {Vanderbilt}(1994)}]{King-Smith:V:1994}%
  \BibitemOpen
  \bibfield  {author} {\bibinfo {author} {\bibfnamefont {R.~D.}\ \bibnamefont
  {King-Smith}}\ and\ \bibinfo {author} {\bibfnamefont {D.}~\bibnamefont
  {Vanderbilt}},\ }\href@noop {} {\bibfield  {journal} {\bibinfo  {journal}
  {Phys. Rev. B}\ }\textbf {\bibinfo {volume} {49}},\ \bibinfo {pages} {5828}
  (\bibinfo {year} {1994})}\BibitemShut {NoStop}%
\bibitem [{\citenamefont {Zhong}\ \emph {et~al.}(1995)\citenamefont {Zhong},
  \citenamefont {Vanderbilt},\ and\ \citenamefont
  {Rabe}}]{Zhong:V:R:PRB:v52:p6301:1995}%
  \BibitemOpen
  \bibfield  {author} {\bibinfo {author} {\bibfnamefont {W.}~\bibnamefont
  {Zhong}}, \bibinfo {author} {\bibfnamefont {D.}~\bibnamefont {Vanderbilt}}, \
  and\ \bibinfo {author} {\bibfnamefont {K.~M.}\ \bibnamefont {Rabe}},\
  }\href@noop {} {\bibfield  {journal} {\bibinfo  {journal} {Phys. Rev. B}\
  }\textbf {\bibinfo {volume} {52}},\ \bibinfo {pages} {6301} (\bibinfo {year}
  {1995})}\BibitemShut {NoStop}%
\bibitem [{\citenamefont {Nishimatsu}\ \emph {et~al.}(2010)\citenamefont
  {Nishimatsu}, \citenamefont {Iwamoto}, \citenamefont {Kawazoe},\ and\
  \citenamefont {Waghmare}}]{Nishimatsu.PhysRevB.82.134106}%
  \BibitemOpen
  \bibfield  {author} {\bibinfo {author} {\bibfnamefont {T.}~\bibnamefont
  {Nishimatsu}}, \bibinfo {author} {\bibfnamefont {M.}~\bibnamefont {Iwamoto}},
  \bibinfo {author} {\bibfnamefont {Y.}~\bibnamefont {Kawazoe}}, \ and\
  \bibinfo {author} {\bibfnamefont {U.~V.}\ \bibnamefont {Waghmare}},\ }\href
  {\doibase 10.1103/PhysRevB.82.134106} {\bibfield  {journal} {\bibinfo
  {journal} {Phys. Rev. B}\ }\textbf {\bibinfo {volume} {82}},\ \bibinfo
  {pages} {134106} (\bibinfo {year} {2010})}\BibitemShut {NoStop}%
\bibitem [{\citenamefont {Yi}\ and\ \citenamefont {He}(2004)}]{Yi2004135}%
  \BibitemOpen
  \bibfield  {author} {\bibinfo {author} {\bibnamefont {Yi}}\ and\ \bibinfo
  {author} {\bibnamefont {He}},\ }\href {\doibase 10.1016/j.tca.2004.02.008}
  {\bibfield  {journal} {\bibinfo  {journal} {Thermochimica Acta}\ }\textbf
  {\bibinfo {volume} {419}},\ \bibinfo {pages} {135 } (\bibinfo {year}
  {2004})}\BibitemShut {NoStop}%
\end{thebibliography}%

\end{document}